\documentclass[aps,prl,twocolumn,superscriptaddress]{revtex4}

\usepackage[dvips]{graphicx}
\usepackage{color}
\usepackage{bm}
\usepackage{amssymb}
\renewcommand{\vec}[1]{\boldsymbol{#1}}
\usepackage[utf8]{inputenc}
\usepackage{times}

\begin{document}

\title{Revisiting the Kitaev material candidacy of Ir$^{4+}$ double perovskite iridates}

\author{A.A. Aczel}
\email{aczelaa@ornl.gov}
\affiliation{Neutron Scattering Division, Oak Ridge National Laboratory, Oak Ridge, TN 37831, USA}
\affiliation{Department of Physics and Astronomy, University of Tennessee, Knoxville, TN 37996, USA}

\author{J.P. Clancy}
\email{clancyp@mcmaster.ca}
\affiliation{Department of Physics and Astronomy, McMaster University, Hamilton, ON L8S 4M1, Canada}

\author{Q. Chen}
\affiliation{Department of Physics and Astronomy, University of Tennessee, Knoxville, TN 37996, USA}

\author{H.D. Zhou}
\affiliation{Department of Physics and Astronomy, University of Tennessee, Knoxville, TN 37996, USA}

\author{D. Reig-i-Plessis}
\affiliation{Department of Physics, University of Illinois at Urbana-Champaign, Urbana, IL 61801, USA}

\author{G.J. MacDougall}
\affiliation{Department of Physics, University of Illinois at Urbana-Champaign, Urbana, IL 61801, USA}

\author{J.P.C. Ruff}
\affiliation{Cornell High Energy Synchrotron Source, Cornell University, Ithaca, NY 14853, USA}

\author{M.H. Upton}
\affiliation{Advanced Photon Source, Argonne National Laboratory, Lemont, IL 60439, USA}

\author{Z. Islam}
\affiliation{Advanced Photon Source, Argonne National Laboratory, Lemont, IL 60439, USA}

\author{T.J. Williams}
\affiliation{Neutron Scattering Division, Oak Ridge National Laboratory, Oak Ridge, TN 37831, USA}

\author{S. Calder}
\affiliation{Neutron Scattering Division, Oak Ridge National Laboratory, Oak Ridge, TN 37831, USA}

\author{J.-Q. Yan}
\affiliation{Materials Science and Technology Division, Oak Ridge National Laboratory, Oak Ridge, TN 37831, USA}

\date{\today}

\begin{abstract}
Quantum magnets with significant bond-directional Ising interactions, so-called Kitaev materials, have attracted tremendous attention recently in the search for exotic spin liquid states. Here we present a comprehensive set of measurements that enables us to investigate the crystal structures, Ir$^{4+}$ single ion properties, and magnetic ground states of the double perovskite iridates La$_2B$IrO$_6$ ($B$~$=$~Mg, Zn) and $A_2$CeIrO$_6$ ($A$~$=$~Ba, Sr) with a large nearest neighbor distance $>$~5~\AA~between Ir$^{4+}$ ions. Our neutron powder diffraction data on Ba$_2$CeIrO$_6$ can be refined in the cubic space group {\it Fm$\bar{3}$m}, while the other three systems are characterized by weak monoclinic structural distortions. Despite the variance in the non-cubic crystal field experienced by the Ir$^{4+}$ ions in these materials, X-ray absorption spectroscopy and resonant inelastic x-ray scattering are consistent with $J_{\rm eff}$~$=$~1/2 moments in all cases. Furthermore, neutron scattering and resonant magnetic x-ray scattering show that these systems host A-type antiferromagnetic order. These electronic and magnetic ground states are consistent with expectations for face-centered-cubic magnets with significant antiferromagnetic Kitaev exchange, which indicates that spacing magnetic ions far apart may be a promising design principle for uncovering additional Kitaev materials.    
\end{abstract}

\maketitle

\section{I. Introduction}

Kitaev materials are quantum magnets with significant nearest neighbour (NN), bond-directional Ising interactions\cite{17_trebst_review, 17_winter_review}. The Kitaev Hamiltonian may be a useful starting point for describing the often exotic magnetic properties of these materials. Notably, for the special cases of the quasi-two-dimensional (quasi-2D) honeycomb lattice\cite{06_kitaev} or its three-dimensional (3D) honeycomb variants\cite{14_lee, 14_kimchi}, these models are exactly soluble and yield a Kitaev spin liquid ground state for either antiferromagnetic (AFM) or ferromagnetic (FM) Kitaev couplings. Numerical approaches or theoretical approximations have been employed to investigate the Kitaev model in other symmetry-allowed cases, revealing a chiral spin liquid\cite{15_li} or nematic phase\cite{15_catuneanu, 15_becker, 16_shinjo} for an AFM Kitaev interaction on the triangular lattice, and an unidentified quantum phase with an extensive degeneracy on the hyperkagome lattice for either AFM or FM Kitaev exchange\cite{14_kimchi_2}. The AFM Kitaev models on the Kagome and pyrochlore lattices are geometrically-frustrated and their ground states are still unknown \cite{14_kimchi_2}. 

Pioneering work by Jackeli and Khaliullin\cite{09_jackeli} provided crucial insights on how to search for new Kitaev materials. Strong electronic correlations and spin-orbit coupling (SOC) can produce $J_{\rm eff}$~$=$~1/2 spin-orbit-assisted Mott insulating states\cite{08_kim} in heavy transition metal (TM) magnets based on Ir$^{4+}$ or Ru$^{3+}$ in an ideal octahedral local environment. The combination of $J_{\rm eff}$~$=$~1/2 single ion wavefunctions and edge-sharing TM-ligand octahedra lead to a complete cancellation of the conventional Heisenberg superexchange via TM-ligand-TM pathways, thus ensuring that the effective magnetic interaction is highly-anisotropic and depends on the spatial orientation of a given bond. 

The quasi-2D honeycomb systems Na$_2$IrO$_3$ \cite{10_singh, 11_jiang}, $\alpha$-Li$_2$IrO$_3$\cite{12_singh, 13_cao_2}, and $\alpha$-RuCl$_3$\cite{14_plumb, 15_johnson, 15_sears, 16_banerjee}, as well as the 3D honeycomb variants $\beta$-Li$_2$IrO$_3$ \cite{15_takayama, 14_biffin_2} and $\gamma$-Li$_2$IrO$_3$\cite{14_modic, 14_biffin}, have been characterized as potential Kitaev materials. This breakthrough started an ongoing quest to search for a Kitaev spin liquid in the laboratory, with the hope of identifying and characterizing the elusive Majorana fermion quasiparticles associated with this state for possible applications in quantum computing. Although it is now well-known that these materials host ordered magnetic ground states due to additional competing interactions, including NN symmetric off-diagonal exchange $\Gamma$, NN Heisenberg direct exchange, and further-neighbor Heisenberg exchange, there have been some promising developments more recently. Inelastic neutron scattering\cite{16_banerjee, 17_banerjee} and Raman spectroscopy measurements\cite{15_sandilands} have shown that $\alpha$-RuCl$_3$ is proximate to the desired Kitaev spin liquid state, as a broad continuum of magnetic scattering has been identified and attributed to fractionalized Majorana fermion excitations. Complementary studies on $\alpha$-RuCl$_3$ have shown that external perturbations, including a magnetic field\cite{18_banerjee}, chemical doping\cite{17_lampen}, and pressure\cite{17_cui} can suppress the zigzag magnetic order and therefore provide a viable way to tune the magnetic Hamiltonian. Additional quasi-2D honeycomb iridates have also been discovered, including Cu$_2$IrO$_3$\cite{17_abramchuk} and H$_3$LiIr$_2$O$_6$\cite{18_kitagawa}, and the initial characterization work has identified dynamical quantum disordered ground states\cite{18_kenney, 18_choi, 18_kitagawa}. While these results are not inconsistent with Kitaev spin liquid physics, significant structural disorder may complicate the picture in both compounds and therefore needs to be better understood. 

\begin{figure}
\centering
\scalebox{0.3}{\includegraphics{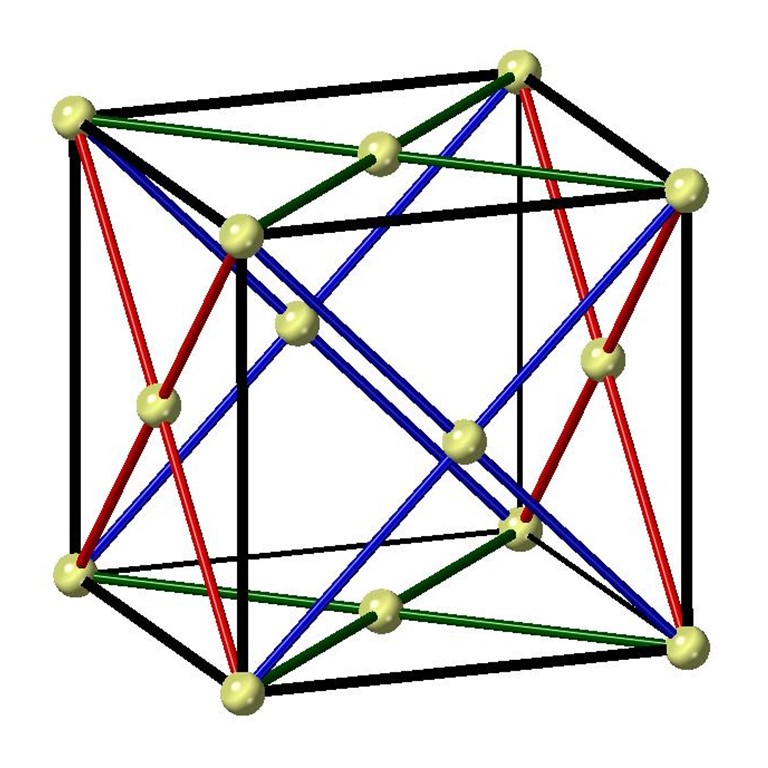}}
\caption{\label{Fig1} (color online) A schematic of the ideal FCC crystal structure showing how NN Kitaev exchange can arise in this case. The colored bonds correspond to bond-directional Ising interactions between the different spin components $S_x$, $S_y$, or $S_z$. For simplicity, the bonds going through the cube are not shown.}
\end{figure}

Recent work on the ideal 6H-perovskite structure, originally thought to be relevant to Ba$_3$IrTi$_2$O$_9$\cite{15_becker}, has shown that materials with adjacent TM-ligand octahedra with parallel edges may also be capable of hosting significant Kitaev interactions. This local geometry for the Ir$^{4+}$ ions ensures that they are coupled through two extended superexchange paths of the form TM-ligand-ligand-TM, possibly leading to a significantly-reduced value for the NN Heisenberg superexchange. It is also tantalizing that spacing the magnetic atoms further apart may lead to the suppression of direct exchange contributions from NN Heisenberg and off-diagonal $\Gamma$ terms, possibly leading to the discovery of Kitaev materials in other structures beyond the honeycomb lattice. While significant Ir/Ti site disorder ensures that Ba$_3$IrTi$_2$O$_9$\cite{12_dey, 17_lee} does not crystallize in the ideal 6H-perovskite structure required for the triangular lattice Kitaev model, the ideas presented in Ref.~\cite{15_becker} led to detailed studies of the double perovskite iridates La$_2$MgIrO$_6$ and La$_2$ZnIrO$_6$ in the context of the face-centered-cubic (FCC) Kitaev model\cite{13_cao, 15_cook, 16_aczel}. Adjacent IrO$_6$ octahedra share parallel edges in the ideal FCC structure and therefore significant NN Kitaev exchange is possible according to the schematic presented in Fig.~\ref{Fig1}. Despite small monoclinic structural distortions leading to non-cubic crystal fields at the Ir$^{4+}$ sites\cite{96_battle, 13_cao}, the classical phase diagram for the FCC Heisenberg-Kitaev-$\Gamma$ model with $J_{\rm eff}$~$=$~1/2 moments is consistent with the A-type AFM ordered states\cite{15_cook} observed in these compounds. Similarly, a magnetic Hamiltonian with a dominant NN AFM Kitaev interaction can explain the Weiss temperatures, the AFM ordering temperatures ($T_N$~$=$~12~K and 7.5 K for the Mg and Zn systems respectively\cite{13_cao}), and the dynamical structure factors measured with inelastic neutron scattering\cite{16_aczel}. 

Ba$_2$CeIrO$_6$ and Sr$_2$CeIrO$_6$ are two other double perovskites with an Ir$^{4+}$ valence inferred from x-ray diffraction measurements\cite{99_wakeshima, 99_harada}, and therefore they are also promising candidates for FCC Kitaev materials based on extended superexchange interactions. Both systems were found to exhibit long-range AFM order, with $T_N$~$=$~17~K and 21~K for the Ba\cite{99_wakeshima} and Sr\cite{99_harada} analogs respectively, although the magnetic structures have yet to be determined. Surprisingly, Ba$_2$CeIrO$_6$ has not been discussed in the context of $J_{\rm eff}$~$=$~1/2 magnetism. On the other hand, two different DFT studies have investigated the $J_{\rm eff}$~$=$~1/2 scenario for Sr$_2$CeIrO$_6$\cite{13_panda, 16_karungo}. While both studies find that electronic correlations and SOC are key ingredients for establishing an AFM insulator, Ref.~\cite{13_panda} supports $J_{\rm eff}$~$=$~1/2 magnetism while Ref.~\cite{16_karungo} argues that significant c-axis compression of the IrO$_6$ octahedra in Sr$_2$CeIrO$_6$ leads to a breakdown of this state and promotes "weak" orbital ordering instead. 

Due to the nearly cubic crystal field environment for Ir$^{4+}$ in the four double perovskite iridates described above, it would be surprising to find significant deviations from $J_{\rm eff}$~$=$~1/2 magnetism in any of these materials, but there are inconsistencies in the published crystal structures (e.g. Ba$_2$CeIrO$_6$ has been reported to crystallize in both cubic\cite{95_ramos} and monoclinic\cite{99_wakeshima} space groups) and direct experimental proof of the Ir$^{4+}$ $J_{\rm eff}$~$=$~1/2 electronic ground states is lacking. Also, there is only limited information available on the collective magnetic properties of these interesting FCC Kitaev material candidates. Therefore, we have revisited their low-temperature crystal structures with neutron powder diffraction (NPD) and assessed their $J_{\rm eff}$~$=$~1/2 candidacy using a combination of x-ray absorption spectroscopy (XAS) and resonant inelastic x-ray scattering (RIXS). We also used scattering techniques to determine the magnetic structures of La$_2$MgIrO$_6$, Ba$_2$CeIrO$_6$, and Sr$_2$CeIrO$_6$. Finally, we performed an inelastic neutron scattering experiment on polycrystalline samples of Ba$_2$CeIrO$_6$ and Sr$_2$CeIrO$_6$, using the same experimental configuration previously reported for La$_2$MgIrO$_6$ and La$_2$ZnIrO$_6$\cite{16_aczel}, in an effort to gain insight into the evolution of the spin waves and hence the magnetic Hamiltonians of the Ce samples relative to the La analogs.  

\section{II. Experimental details}

Polycrystalline samples of La$_2$MgIrO$_6$, La$_2$ZnIrO$_6$, Ba$_2$CeIrO$_6$ and Sr$_2$CeIrO$_6$ were synthesized by conventional solid state reactions. The detailed procedure for synthesizing La$_2$MgIrO$_6$ and La$_2$ZnIrO$_6$ is presented in Ref.~\cite{13_cao}. For Ba$_2$CeIrO$_6$ and Sr$_2$CeIrO$_6$, the starting materials \textit{A}CO$_3$ (\textit{A}~$=$~Ba, Sr), CeO$_2$, and IrO$_2$ were first mixed in the appropriate molar ratio. The homogeneous powder was then pelletized, placed in a covered alumina crucible, and heated to 1200$^\circ$C in 20 hours and held at this temperature for 60 hrs. The heating process was repeated again after one intermediate grinding. X-ray powder diffraction measurements verified that all four polycrystalline samples were single phase and magnetic susceptibility measurements using a Quantum Design Magnetic Property Measurement System (MPMS) confirmed the AFM ordering temperatures reported previously\cite{13_cao, 99_wakeshima, 99_harada}. 

Single crystals of La$_2$MgIrO$_6$ were grown by a flux method using a stoichiometric amount of the starting materials La$_2$O$_3$, MgO and Ir with purities not less than 99.9\%. A mixture of PbO-PbF$_2$ with a molar ratio of 1:1 was used as the flux by combining it with the starting materials in a 6:1 mass ratio. The mixed powder was placed into a covered platinum crucible and heated to 1220$^\circ$C at 120$^\circ$C/hr, left at the target temperature for 24 hours, and then cooled down to 500$^\circ$C at 15$^\circ$C/hr. The furnace was then turned off and the crystals were separated from the flux with a centrifuge. They typically had an octahedral geometry and all dimensions were less than 1 mm. X-ray powder diffraction measurements on crushed single crystals verified phase purity and magnetic susceptibility measurements using a Quantum Design MPMS confirmed that the ordering temperature was consistent with both previous and current results on polycrystalline samples. Unfortunately, attempts to grow single crystals of the other three compositions using a similar flux procedure were unsuccessful. 

Neutron powder diffraction (NPD) was performed with $\sim$~5 g of polycrystalline La$_2$MgIrO$_6$, La$_2$ZnIrO$_6$, and Ba$_2$CeIrO$_6$ and $\sim$~2.5~g of polycrystalline Sr$_2$CeIrO$_6$ using the HB-2A powder diffractometer of the High Flux Isotope Reactor (HFIR) at Oak Ridge National Laboratory (ORNL) to revisit the crystal structures of these materials systematically. The samples were loaded in cylindrical vanadium cans with 5~mm inner diameters. The data was collected at $T$~$=$~4~K with a neutron wavelength of 1.54~\AA~and slightly different collimations of 12$'$-21$'$-6$'$ for La$_2$MgIrO$_6$ and La$_2$ZnIrO$_6$ and open-21$'$-12$'$ for Ba$_2$CeIrO$_6$ and Sr$_2$CeIrO$_6$. 

X-ray absorption spectroscopy (XAS) was performed on polycrystalline samples of La$_2$MgIrO$_6$, La$_2$ZnIrO$_6$, Ba$_2$CeIrO$_6$ and Sr$_2$CeIrO$_6$ at room temperature using the A2 beamline at the Cornell High Energy Synchrotron Source (CHESS) to assess the importance of spin-orbit coupling to their Ir$^{4+}$ electronic ground states. Measurements were collected at both the $L_2$ (2$p_{1/2}$~$\rightarrow$~5$d$) and $L_3$  (2$p_{3/2}$~$\rightarrow$~5$d$) Ir absorption edges, which occur at energies of 12.824~keV and 11.215~keV respectively. The energy of the incident x-ray beam was selected using a diamond-(1 1 1) double crystal monochromator, with higher harmonic contributions suppressed by a combination of Rh-coated mirrors and a 50\% detuning of the second monochromator crystal. The XAS measurements were performed in transmission geometry, using a series of three ion chambers ($I_0$, $I_1$, and $I_2$). The sample was mounted between $I_0$ and $I_1$, while an elemental Ir reference sample was mounted between $I_1$ and $I_2$. This configuration allows a direct measurement of the linear x-ray attenuation coefficient, $\mu(E)$, which is defined by the intensity ratio of the incident and transmitted x-ray beams. In this case, $\mu_{sample}(E)$~$=$~$I_0/I_1$ and $\mu_{Ir}(E)$~$=$~$I_1/I_2$. The energy calibration of this setup is accurate to within 0.25 eV, and direct comparisons between sample and reference spectra can be used to rule out any systematic energy drifts over the course of the experiment. 

Resonant inelastic x-ray scattering (RIXS) measurements were conducted on polycrystalline samples of La$_2$MgIrO$_6$, La$_2$ZnIrO$_6$, Ba$_2$CeIrO$_6$ and Sr$_2$CeIrO$_6$ at room temperature using the MERIX spectrometer on beamline 27-ID of the Advanced Photon Source (APS) at Argonne National Laboratory to investigate the Ir$^{4+}$ crystal field excitations. The incident x-ray energy was tuned to the Ir $L_3$ absorption edge at 11.215 keV. A double-bounce diamond-(1 1 1) primary monochromator, a channel-cut Si-(8 4 4) secondary monochromator, and a spherical (2 m radius) diced Si-(8 4 4) analyzer crystal were used to obtain an overall energy resolution of $\sim$~35 meV (full width at half maximum [FWHM]). In order to minimize the elastic background intensity, measurements were carried out in horizontal scattering geometry with the scattering angle 2$\theta$ set to 90$^\circ$. 

Resonant magnetic x-ray scattering (RMXS) measurements were performed on a single crystal of La$_2$MgIrO$_6$ using beamline 6-ID-B at the APS to determine the magnetic structure of this material. The incident x-ray energy was tuned to the Ir $L_3$ absorption edge at 11.215 keV. Measurements were carried out in vertical scattering geometry, using incident photons which were linearly polarized perpendicular to the scattering plane (i.e. $\sigma$ polarization). In this geometry, resonant magnetic scattering rotates the plane of linear polarization into the scattering plane (i.e. $\pi$ polarization). In contrast, charge scattering does not change the polarization of the scattered photons. As a result, polarization analysis of the scattered beam can be used to distinguish the magnetic ($\sigma$-$\pi$) and charge ($\sigma$-$\sigma$) scattering contributions. The (0 0 8) reflection from pyrolytic graphite (PG) was used as a polarization and energy analyzer. The sample was mounted on the cold finger of a closed-cycle refrigerator capable of reaching temperatures from 6 K to 300 K. We were able to identify a crystal with a surface normal corresponding to the $[1 1 0]$ direction (indexed in {\it P2$_1$/n} monoclinic notation) for this experiment. Our measurements primarily focused on reflections close to this surface normal direction. 

Elastic neutron scattering measurements, complementary to the NPD experiment described above, were performed on the 14.6~meV fixed-incident-energy triple-axis spectrometer HB-1A of the HFIR at ORNL using the same polycrystalline samples of Ba$_2$CeIrO$_6$ and Sr$_2$CeIrO$_6$ studied at HB-2A. Since the main goal of this experiment was to determine the magnetic structures of these materials, these samples were loaded in Al cans with a 1 mm thick annulus to minimize neutron absorption. The background was also minimized by using a double-bounce monochromator system, mounting two-highly oriented PG filters in the incident beam to remove higher-order wavelength contamination, and placing a PG analyzer crystal before the single He-3 detector for energy discrimination. A collimation of 40$'$-40$'$-80$'$-open resulted in an energy resolution at the elastic line just over 1 meV (FWHM). The elastic scattering was measured at 4~K and 30~K for both samples. 

Inelastic neutron scattering (INS) spectra were measured with the HYSPEC spectrometer at the Spallation Neutron Source of ORNL using $\sim$~5~g of polycrystalline Ba$_2$CeIrO$_6$ and Sr$_2$CeIrO$_6$ loaded into the annular cans described above. All data were collected using incident energies of $E_i$~$=$~7.5 and 15 meV, with corresponding Fermi chopper frequencies of 240 and 300~Hz, resulting in instrumental energy resolutions of 0.3 and 0.7 meV (FWHM) respectively at the elastic line. A He cryostat was used to achieve a base temperature of 1.5~K. Empty Al annular can measurements were subtracted from all the HYSPEC data presented in this work to minimize the Al scattering contribution to the sample spectra.

\section{III. Crystal structures}

$B/B'$-site ordered double perovskites, with the general formula $A_2 B B' $O$_6$, may crystallize in an ideal FCC structure or lower-symmetry variants. Detailed knowledge of the Ir$^{4+}$ local environment is crucial for properly assessing the $J_{\rm eff}$~$=$~1/2 candidacy of the double perovskite iridates La$_2$MgIrO$_6$, La$_2$ZnIrO$_6$, Ba$_2$CeIrO$_6$ and Sr$_2$CeIrO$_6$, as significant non-cubic crystal fields at the Ir$^{4+}$ sites can lead to deviations in the electronic wavefunctions expected for this desired single ion ground state. It is also important to characterize symmetry-lowering structural distortions away from an ideal FCC lattice since they can lead to additional NN interactions between Ir ions beyond those captured in the Heisenberg-Kitaev-$\Gamma$ model (e.g. anisotropic Kitaev exchange\cite{16_aczel}). The crystal structures have been determined previously by room temperature x-ray powder diffraction\cite{95_currie, 95_ramos, 99_harada, 99_wakeshima,13_cao} and variable temperature neutron powder diffraction\cite{96_battle, 00_harada, 16_karungo}. Most refinements show that these materials exhibit very small monoclinic distortions away from the ideal FCC structure, although Ba$_2$CeIrO$_6$ has been reported to crystallize in both cubic\cite{95_ramos} and monoclinic\cite{99_wakeshima} space groups. Some variation has also been found in the structural parameters crucial for establishing $J_{\rm eff}$~$=$~1/2 electronic ground states in these materials, such as the oxygen fractional coordinates, Ir-O bond lengths, and O-Ir-O bond angles. X-ray diffraction is not the optimal technique for determining these parameters due to the weak x-ray scattering power of oxygen, while the specific sample geometry used in a NPD experiment affects the relative intensities of the Bragg peaks due to the significant neutron absorption expected from iridium and this introduces systematic error into the refinements. Some previous diffraction work has also assumed no $B/B'$-site mixing\cite{96_battle, 00_harada}, even though it is a common feature of this crystal structure. Therefore, an NPD study where the crystal structures of these materials are revisited with a consistent sample geometry is highly-warranted so the results can be compared on equal footing. 

\begin{figure}
\centering
\scalebox{0.37}{\includegraphics{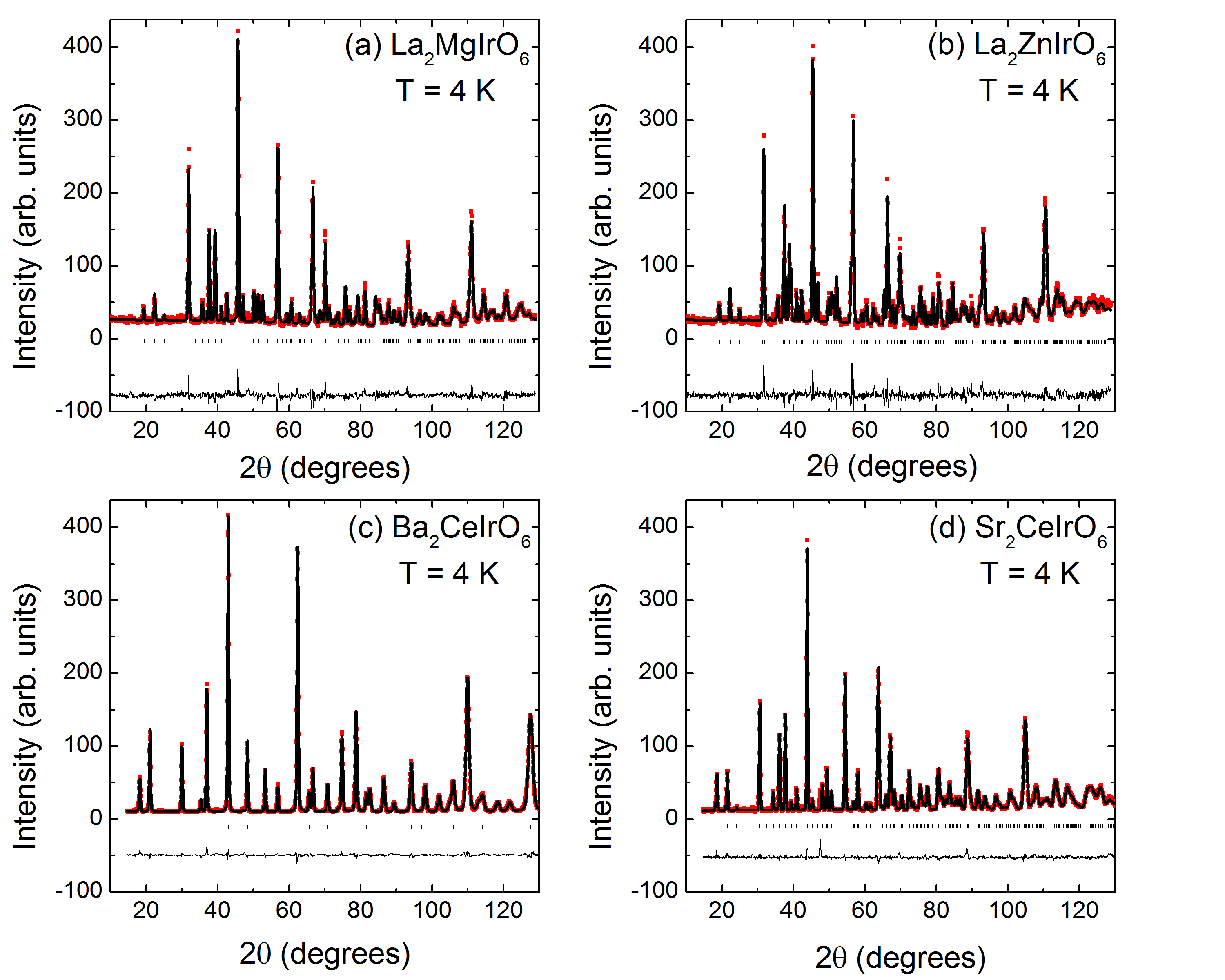}}
\caption{\label{Fig2} (color online) Neutron powder diffraction data, indicated by the solid symbols and collected with a neutron wavelength~1.54~\AA~at a temperature $T$~$=$~4~K, is shown for (a) La$_2$MgIrO$_6$, (b) La$_2$ZnIrO$_6$, (c) Ba$_2$CeIrO$_6$ and (d) Sr$_2$CeIrO$_6$. The best structural refinements are superimposed on the data as solid curves, the difference curves are shown below the diffraction patterns, and the expected Bragg peak positions are indicated by ticks. }
\end{figure}

\begin{table}[htb]
\begin{center}
\caption{Structural parameters for La$_2B$IrO$_6$ ($B$~$=$~Mg, Zn) and $A_2$CeIrO$_6$ ($A$~$=$~Ba, Sr) extracted from the refinements of the~1.54~\AA~neutron powder diffraction data. The lattice constants and bond distances are in \AA~and all angles are in degrees.} 

\begin{tabular}{l l l l l}
\hline 
\hline
Material & La-Mg & La-Zn & Ba-Ce & Sr-Ce \\
\hline
Space Group & {\it P$2_1$/n} & {\it P$2_1$/n} & {\it Fm$\bar{3}$m} & {\it P$2_1$/n} \\
$a$ & 5.5874(2) & 5.5917(2) & 8.4126(1) & 5.8243(2) \\  
$b$ & 5.6307(2) & 5.6912(2) & 8.4126(1) & 5.8400(2) \\
$c$ & 7.9119(3) & 7.9335(3) & 8.4126(1) & 8.2395(3) \\
$\beta$ & 90.01(1) & 90.03(1) & 90  & 90.266(3) \\
A $x$ & 0.512(1) & 0.5112(9) & 0.25 & 0.5074(8) \\ 
A $y$ & 0.5411(5) & 0.5497(5) & 0.25 & 0.5336(4) \\ 
A $z$ & 0.253(2) & 0.248(1) & 0.25     & 0.2460(8)             \\
B & (0.5,0,0) & (0.5,0,0) & (0.5,0.5,0.5) & (0.5,0,0.5)   \\
Ir & (0.5,0,0.5) & (0.5,0,0.5) & (0,0,0) & (0.5,0,0)  \\
O$_1$ $x$ & 0.215(2) & 0.200(2) & 0.2390(1) & 0.226(1) \\
O$_1$ $y$ & 0.221(2) & 0.214(2) & 0 & 0.2053(9) \\
O$_1$ $z$ & 0.951(1) & 0.945(1) & 0 & 0.9636(7) \\
O$_2$ $x$ & 0.295(2) & 0.293(2) & 0.2390(1) & 0.3005(9) \\
O$_2$ $y$ & 0.699(2) & 0.699(2) & 0 & 0.726(1) \\
O$_2$ $z$ & 0.964(1) & 0.963(1) & 0 & 0.9598(6) \\
O$_3$ $x$ & 0.420(1) & 0.410(1) & 0.2390(1) & 0.4259(9) \\
O$_3$ $y$ & 0.986(1) & 0.980(1) & 0 & 0.9847(7) \\
O$_3$ $z$ & 0.256(2) & 0.254(2) & 0 & 0.2368(6) \\
Site mixing & 20(3)~\% & 14(3)~\% & 0 & 0 \\
R$_\mathrm{wp}$ & 8.15~\% & 9.65~\% & 4.40~\% & 6.38~\% \\ 
$\chi^2$ & 1.67 & 2.28 & 6.32 & 9.20 \\
Ir-O$_1$ & 2.02(1) & 2.02(1) & 2.010(1) & 2.018(6) \\
Ir-O$_2$ & 2.01(1) & 2.01(1) & 2.010(1) & 2.007(6) \\ 
Ir-O$_3$ & 1.98(2) & 2.02(2) & 2.010(1) & 2.002(5) \\ 
O$_1$-Ir-O$_2$ &  94.7(8)          &  91.6(8)         & 90          & 90.4(4)     \\ 
O$_2$-Ir-O$_3$ &  91.3(8)          &  91.7(8)          & 90           &  90.1(4)       \\
O$_1$-Ir-O$_3$ &  91.3(9)          &  91.5(9)         & 90           &   90.2(4)     \\
$w(Ir)$ & 9 & 11 & - & 9 \\
$j(Ir)$ & 13 & 14 & - & 12   \\
\hline\hline
\end{tabular}
\end{center}
\end{table}

Figure~\ref{Fig2} shows HB-2A NPD data as solid red squares collected using a neutron wavelength of~1.54~\AA~for all four double perovskite iridates with $T$~$=$~4~K. A common sample geometry was used for these measurements with each composition loaded in a cylindrical vanadium can with a 5~mm inner diameter. Rietveld refinement results performed using FullProf \cite{93_rodriguez} are superimposed on the data as black solid curves. Table I shows lattice constants, atomic fractional coordinates, and selected bond distances and angles extracted from the refinements. We find that the Ba$_2$CeIrO$_6$ data refines well in the space group {\it Fm$\bar{3}$m} corresponding to the ideal FCC structure, which implies that the Ir$^{4+}$ ions in this material have cubic point symmetry and hence a $J_{\rm eff}$~$=$~1/2 electronic ground state. In sharp contrast, the diffraction data for the other three materials are better described by the monoclinic {\it P$2_1$/n} space group. Significant $B/B'$ site mixing is found in La$_2$MgIrO$_6$ (20~\%) and La$_2$ZnIrO$_6$ (14~\%), which is larger than past estimates from x-ray diffraction\cite{16_aczel} but agrees well with NPD results on the related materials Sr$_2$MgIrO$_6$ and Sr$_2$ZnIrO$_6$\cite{15_laguna}. No $B/B'$-site mixing is found in Ba$_2$CeIrO$_6$ or Sr$_2$CeIrO$_6$. The amount of $B/B'$ site mixing inversely tracks the $B/B'$ ionic radii (r) differences in these materials, as r$_{Ce^{4+}}$~$>$~r$_{Zn^{2+}}$~$>$~r$_{Mg^{2+}}$~$>$~r$_{Ir^{4+}}$\cite{76_shannon}. No extra Bragg peaks indicative of magnetic order are observed in this data, likely due to extremely small ordered moments. 

Typically, structural distortions arise in $B/B'$-site ordered double perovskites due to a small $A$-site cation, with the ideal {\it Fm$\bar{3}$m} FCC structure often becoming tetragonal {\it I4/m} or monoclinic {\it P2$_1$/n}. Table I shows that the double perovskite iridates studied here follow this general trend, as the structural distortions become larger when the ionic radius of the $A$-site decreases (i.e. r$_{Ba^{2+}}$~$>$~r$_{Sr^{2+}}$~$>$~r$_{La^{3+}}$)\cite{76_shannon}. Assuming that $\hat{x}$, $\hat{y}$, and $\hat{z}$ are aligned with the three FCC crystallographic directions, the relationships between the tetragonal and FCC lattice vectors are as follows: $\vec{a}_{\rm t}$~$=$~$a_{\rm FCC}(\hat{x}\pm\hat{y})/2$ and $\vec{c}_{\rm t}$~$=$~$a_{\rm FCC} \hat{z}$. The monoclinic structure is then derived from the tetragonal unit cell by simply modifying the lattice constants such that $\vec{a} \neq \vec{b}$ and $\beta \neq 90^\circ$. The $B'$O$_6$ octahedra are both distorted and rotated as a consequence of the symmetry-lowering. 

The non-cubic crystal fields $\Delta$ at the Ir$^{4+}$ sites in the monoclinic double perovskite iridates, which may lead to possible deviations from $J_{\rm eff}$~$=$~1/2 magnetism, arise from IrO$_6$ octahedral distortions only. The relative magnitudes and signs of $\Delta$ can be determined by comparing Ir-O bond lengths and O-Ir-O bond angles associated with these octahedra. Our refinement results reported in Table I show that all six Ir-O bond lengths for the three monoclinic double perovskite iridates are within 2\%~of each other, while all O-Ir-O bond angles are within 4.7$^\circ$ (La$_2$MgIrO$_6$), 1.7$^\circ$ (La$_2$ZnIrO$_6$), or 0.4$^\circ$ (Sr$_2$CeIrO$_6$) of the ideal 90$^\circ$ and 180$^\circ$ values. These deviations from an Ir$^{4+}$ cubic crystal field environment are smaller or comparable to the IrO$_6$ octahedral distortions measured in the $J_{\rm eff}$~$=$~1/2 magnets Na$_2$IrO$_3$\cite{12_ye} and Sr$_2$IrO$_4$\cite{94_crawford}, and therefore these diffraction results provide indirect evidence that the $J_{\rm eff}$~$=$~1/2 description applies to the monoclinic double perovskite iridates described here. 

The monoclinic structure also ensures that adjacent IrO$_6$ octahedra no longer have parallel edges, which can have a significant effect on the exchange interactions. Both the IrO$_6$ octahedral distortions and rotations described above can play a role here. The IrO$_6$ octahedral distortions have already been quantified in La$_2$MgIrO$_6$, La$_2$ZnIrO$_6$, and Sr$_2$CeIrO$_6$ as explained above, while the octahedral rotations can be determined according to Ref.~\cite{86_groen} by using the refined atomic fractional coordinates and the Glazer notation discussed in Refs.~\cite{72_glazer, 97_woodward}. We find that the IrO$_6$ octahedra are subjected to a global rotation $j(Ir)$ about the b-axis and another rotation $w(Ir)$ about the c-axis that is staggered between adjacent ab-layers; the magnitude of these rotations is shown in Table I. Overall, our results indicate that the monoclinic double perovskite iridates are characterized by very weak IrO$_6$ octahedral distortions but significant IrO$_6$ octahedral rotations, and therefore the latter effect will lead to the largest deviations in the collective magnetic properties expected for ideal FCC systems. 

\section{IV. Single ion properties}

Despite the perceived importance of spin-orbit coupling effects in La$_2$MgIrO$_6$, La$_2$ZnIrO$_6$, Ba$_2$CeIrO$_6$ and Sr$_2$CeIrO$_6$, no attempt has been made to measure the strength of the spin-orbit interactions in these compounds. One established technique for accomplishing this task is x-ray absorption spectroscopy (XAS), as it was shown that the integrated intensity ratio of the white line features measured at the $L_2$ and $L_3$ absorption edges (i.e. the branching ratio $BR$~$=$~$I_{L_3}/I_{L_2}$) is directly proportional to the expectation value of the spin-orbit coupling operator $<\vec{L} \cdot \vec{S}>$\cite{88_laan, 88_thole_1, 88_thole_2}. More specifically, the branching ratio can be written as $BR$~$=$~$(2+r)/(1-r)$, where $r$~$=$~$<\vec{L} \cdot \vec{S}>/<n_h>$ and $n_h$ is the number of holes in the valence shell. A branching ratio significantly greater than 2 indicates a strong coupling between the local orbital and spin moments in the electronic ground state of the transition metal under investigation. However, the converse is not necessarily true (i.e. a statistical BR can arise even in the presence of strong spin-orbit coupling if the electronic bandwidth is sufficiently large or the spin or orbital moments are quenched).

\begin{figure}
\centering
\scalebox{0.22}{\includegraphics{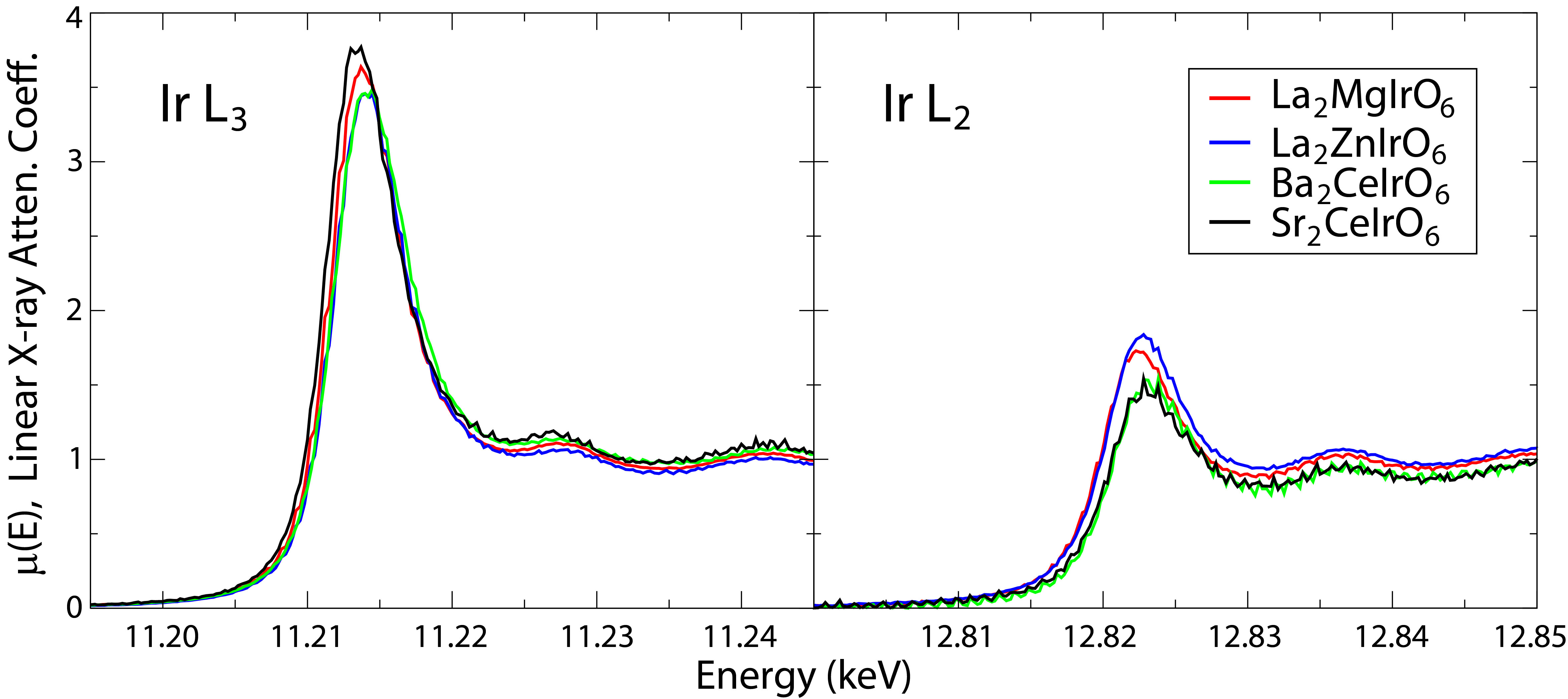}}
\caption{\label{Fig3} (color online) X-ray absorption spectra collected at the Ir $L_3$ edge (left) and Ir $L_2$ edge (right) for the double perovskite iridates La$_2$MgIrO$_6$, La$_2$ZnIrO$_6$, Ba$_2$CeIrO$_6$ and Sr$_2$CeIrO$_6$. Note the dramatic intensity difference between the sharp ``white-line'' features observed at the $L_3$ and $L_2$ absorption edges.}
\end{figure}

Fig.~\ref{Fig3} shows the x-ray absorption spectra at the Ir $L_3$ and $L_2$ edges, plotted as the linear x-ray attenuation coefficient $\mu(E)$ vs energy, for all four double perovskite iridates. The data was normalized to absorption steps of 1 for both the $L_3$ and $L_2$ edges. To extract the BR from our data, we followed the procedure described in Ref.~\cite{12_clancy}. More specifically, we fit the near edge portion of each spectrum to the following expression:
\begin{equation}
\mu(E) = C_0 + C_1E + C_2 \rm{arctan}\left(\frac{E-E_0}{\Gamma/2}\right) + \frac{C_3}{1+\left(\frac{E-E_0}{\Gamma/2}\right)^2} 
\end{equation}
where $C_0$ and $C_1$ represent a linear background, $C_2$ is the absorption step height, $C_3$ is the white-line intensity, and $E_0$ and $\Gamma$ correspond to the center and width of both the arctangent and Lorentzian functions. A secondary measure of the white-line intensity was also obtained from simple numerical integration, after modeling the absorption step by a unit step function fixed at the inflection point of $\mu$(E). This numerical integration provides a useful consistency check for the fit results, as well as a measure of white-line intensity that is less sensitive to any lineshape asymmetry. The final values for the white-line intensities, and their corresponding BRs, reflect the average obtained from these two methods. The BRs and $<\vec{L} \cdot \vec{S}>$ values extracted from this analysis are shown in Table~II, where they are also compared to the values obtained from other selected $d^{5}$ iridates. The enhanced BRs for La$_2$MgIrO$_6$, La$_2$ZnIrO$_6$, Ba$_2$CeIrO$_6$ and Sr$_2$CeIrO$_6$ are within the range typically found for an Ir$^{4+}$ ion in an octahedral local environment. These results indicate sizable orbital contributions to the Ir$^{4+}$ electronic ground states in these materials, which is an essential ingredient for the realization of $J_{\rm eff}$~$=$~1/2 moments. 

\begin{table}[htb]
\begin{center}
\caption{Branching ratios (BR) and expectation value of the spin-orbit coupling operator $<\vec{L} \cdot \vec{S}>$ in units of $\hbar^2$ for selected $d^5$ iridates.} 
\begin{tabular}{l l l l}
\hline 
\hline
Material & BR & $<\vec{L} \cdot \vec{S}>$  & Ref. \\
\hline
La$_2$MgIrO$_6$ & 6.5(9) & 3.0(6) & this work \\  
La$_2$ZnIrO$_6$ & 4.5(5) & 2.3(4) & this work \\
Ba$_2$CeIrO$_6$ & 6.8(9) & 3.1(6) & this work \\
Sr$_2$CeIrO$_6$ & 6.3(7) & 3.0(5) & this work \\
Sr$_2$TiIrO$_6$ & 4.04 & 2.02 & \cite{15_laguna} \\
La$_2$NiIrO$_6$ & 4.31 & 2.18 & \cite{15_laguna} \\
Sr$_2$IrO$_4$ & 7.0(4) & 3.0(3) & \cite{12_clancy} \\ 
Sr$_3$Ir$_2$O$_7$ & 5.5 & 2.69 & \cite{18_donnerer}   \\
Na$_2$IrO$_3$ & 5.7(3) & 2.8(2) & \cite{12_clancy} \\ 
$\alpha$-Li$_2$IrO$_3$ & 5.1(4) & 2.5(2) & \cite{18_clancy} \\
Y$_2$Ir$_2$O$_7$ & 6.0(3) & 2.9(2) &  \cite{12_clancy} \\
\hline\hline
\end{tabular}
\end{center}
\end{table}

Unfortunately, establishing $J_{\rm eff}$~$=$~1/2 magnetism in $d^5$ iridates on the basis of BR measurements alone has proven to be extremely difficult. When the SOC coupling constant $\lambda$~$<<$~the cubic crystal field splitting 10$Dq$, one expects $<\vec{L} \cdot \vec{S}>$~$=$~$\hbar^2$ for a $J_{\rm eff}$~$=$~1/2 state\cite{10_laguna}. The $<\vec{L} \cdot \vec{S}>$ values in Table II are significantly larger, even for established $J_{\rm eff}$~$=$~1/2 magnets like Sr$_2$IrO$_4$ and Sr$_3$Ir$_2$O$_7$. It is now well-known that $\lambda$~$\sim$~$\frac{1}{6}10Dq$ is typical for $d^5$ iridates\cite{18_donnerer}, and so the assumption made above to calculate the BR for a $J_{\rm eff}$~$=$~1/2 state may not be valid. Instead, significant mixing between the excited $J_{\rm eff}$~$=$~3/2 and $e_g$ manifolds will yield enhanced BRs\cite{10_laguna, 18_stamokostas} with a range of possible material-dependent values even for $J_{\rm eff}$~$=$~1/2 magnets. This being said, the presence of a significantly enhanced BR does appear to be a good indicator of the $J_{\rm eff}$~$=$~1/2 state in several other candidate Kitaev materials.  In particular, the pressure-induced collapse of the $J_{\rm eff}$~$=$~1/2 ground state in $\alpha$-Li$_2$IrO$_3$ is accompanied by a rapid drop in the BR\cite{18_clancy}, which approaches that of elemental Ir (BR $\sim$ 3) in its dimerized, non-$J_{\rm eff}$, high pressure state.  

\begin{figure}
\centering
\scalebox{0.95}{\includegraphics{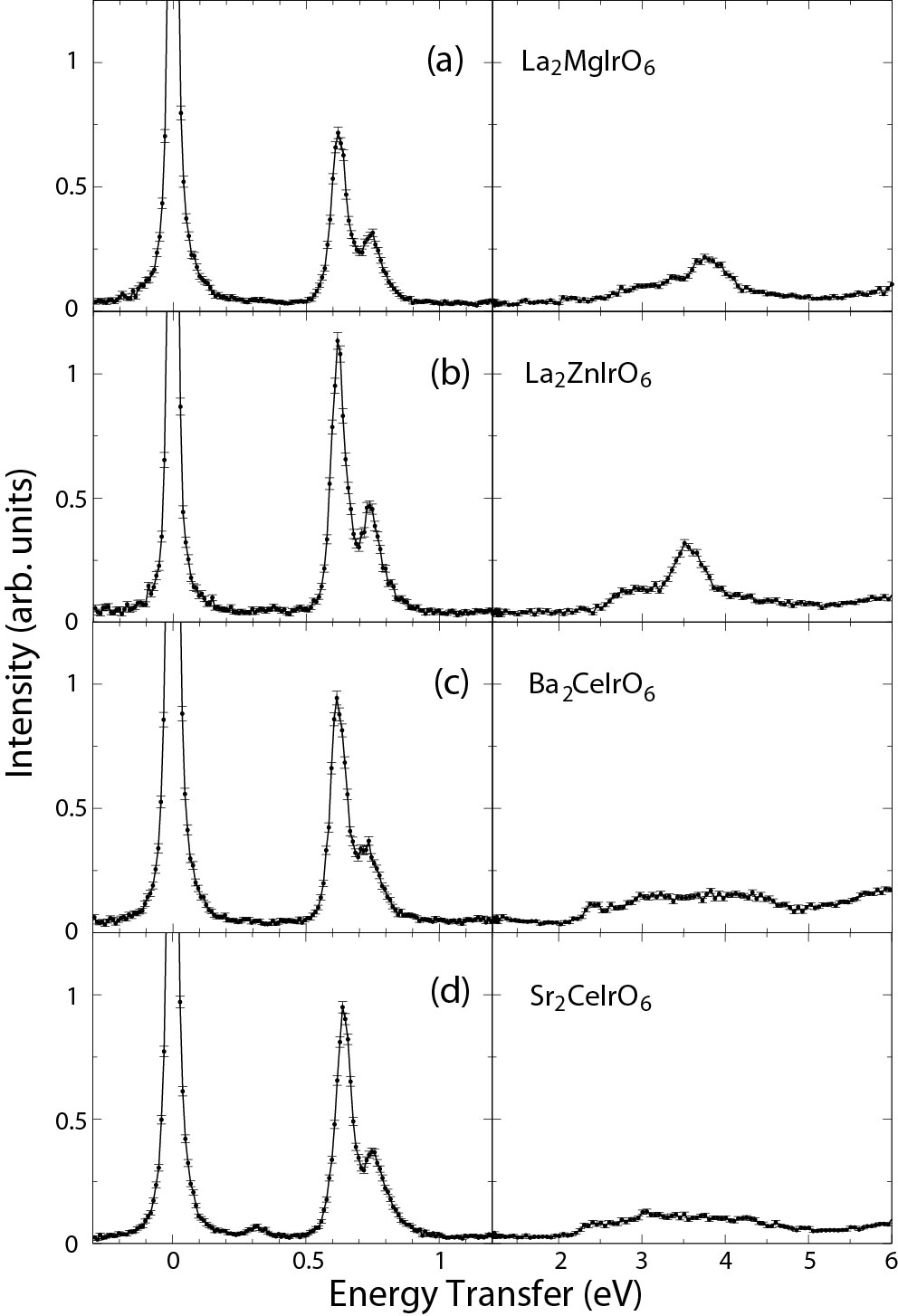}}
\caption{\label{Fig4} (color online) Resonant inelastic x-ray scattering spectra collected at the Ir $L_3$ edge ($E_i$ = 11.215 keV) for (a) La$_2$MgIrO$_6$, (b) La$_2$ZnIrO$_6$, (c) Ba$_2$CeIrO$_6$ and (d) Sr$_2$CeIrO$_6$. Note the presence of a strong elastic line at an energy transfer $\hbar \omega$~$=$~0, two peaks at energy transfers of $\hbar \omega $~$\sim$ 0.6 to 0.7 eV corresponding to intraband $t_{2g}$ crystal field transitions, and higher energy excitations centered at $\hbar \omega $~$\sim$ 3.5 eV corresponding to interband $t_{2g}$ to $e_{g}$ crystal field transitions.}
\end{figure}

\begin{table}[htb]
\begin{center}
\caption{RIXS fitting results of the intra-$t_{2g}$ excitations, spin-orbit coupling constants ($\lambda$), and non-cubic crystal field splitting ($\Delta$) of the J$_{\rm eff}$~$=$~3/2 manifold for selected iridates and iridium fluorides. All parameters are in meV.} 
\begin{tabular}{l l l l l l}
\hline 
\hline
Material & $\hbar \omega_1$ & $\hbar \omega_2$ & $\lambda$ & $\Delta$ & Ref. \\
\hline
La$_2$MgIrO$_6$ & 625(1) & 744(2) & 456(1) & 119(2) & this work \\  
La$_2$ZnIrO$_6$ & 624(1) & 744(2) & 456(1) & 120(2) & this work \\
Ba$_2$CeIrO$_6$ & 625(1) & 735(4) & 453(1) & 110(4) & this work \\
Sr$_2$CeIrO$_6$ & 645(1) & 760(3) & 468(1) & 115(3) & this work \\
Sr$_2$IrO$_4$ & 550 & 700 & 417 & 150 & \cite{17_lu} \\ 
Sr$_3$Ir$_2$O$_7$ & 500 & 700 &  400  &  200  & \cite{17_lu}   \\
Na$_2$IrO$_3$ & 720(20) & 830(20) & 517(9) & 110(30) & \cite{13_gretarsson} \\ 
$\alpha$-Li$_2$IrO$_3$ & 720(20) & 830(20) & 517(9) & 110(30) & \cite{13_gretarsson} \\
Y$_2$Ir$_2$O$_7$ & 530(50) & 980(50) & 500(20) & 450(70) & \cite{14_hozoi} \\
K$_2$IrF$_6$ & 802(1) & 914(1) & 572(1) & 112(1) & \cite{17_rossi} \\
Na$_2$IrF$_6$ & 816(1) & 923(1) & 580(1) & 107(1) & \cite{17_rossi} \\
\hline\hline
\end{tabular}
\end{center}
\end{table}

Resonant inelastic x-ray scattering (RIXS) provides a direct measurement of the electronic ground state and therefore offers a uniquely powerful way to assess the $J_{\rm eff}$~$=$~1/2 candidacy of the $d^5$ iridates. For an Ir$^{4+}$ ion in an ideal octahedral environment, the $J_{\rm eff}$~$=$~1/2 doublet ground state is separated from the $J_{\rm eff}$~$=$~3/2 quartet excited state by $\frac{3}{2}\lambda$. Non-cubic crystal fields at the Ir$^{4+}$ site will split the excited quartet into two doublets. RIXS can be used to probe these crystal field excitations and therefore provides a direct measurement of $\lambda$ and the non-cubic crystal field $\Delta$ at the Ir$^{4+}$ site. 

Fig.~\ref{Fig4} presents the RIXS spectra at the Ir $L_3$ edge for La$_2$MgIrO$_6$, La$_2$ZnIrO$_6$, Ba$_2$CeIrO$_6$ and Sr$_2$CeIrO$_6$; two inelastic features are observed for each sample below 1~eV. The energy scale of these modes matches well with expectations for intraband $t_{2g}$ crystal field transitions in $d^5$ iridates\cite{18_kim}. We fit each spectrum to the sum of three Lorentzian functions representing the elastic line and the two intraband $t_{2g}$ transitions. These fits were used to establish precise inelastic peak positions ($\hbar \omega_1$ and $\hbar \omega_2$) and therefore enable a meaningful quantitative comparison between the four samples; the fitting results are summarized in Table~III. Several past studies on other iridates have extracted $\lambda$ and $\Delta$ from $\hbar \omega_1$ and $\hbar \omega_2$ using a simple single ion Hamiltonian with spin-orbit coupling and tetragonal crystal field terms\cite{12_liu, 14_hozoi, 14_sala}. Since the DP iridates being considered here have a more complicated non-cubic crystal field splitting, we adopt a different approach to estimate $\lambda$ and $\Delta$. We assume that the energy difference between $\hbar \omega_1$ and $\hbar \omega_2$ corresponds to $\Delta$, while the average energy of these two peaks is $\frac{3}{2} \lambda$. We note that this method provides a reasonable estimation of $\lambda$, especially when $\Delta$~$\le$~200~meV\cite{14_sala_2}. 

Our results for $\lambda$ and $\Delta$ are shown in Table~III and compared to the values obtained for other selected iridates and iridium fluorides using this same approach. Sr$_2$CeIrO$_6$ has a slightly larger $\lambda$ value as compared to the other three DP iridates, while $\Delta$ follows the general trend expected from NPD and increases as the Ir$^{4+}$ local environment becomes progressively more distorted. The most surprising finding may be the presence of two intra-$t_{2g}$ excitations in the Ba$_2$CeIrO$_6$ spectrum despite the assignment of {\it Fm$\bar{3}$m} cubic symmetry from NPD, but this may arise due to a small global structural distortion that was not resolved by neutron diffraction or local distortions of the IrO$_6$ octahedra. Nonetheless, these results show that $\lambda/\Delta$~$>$~3.5 for all four double perovskite iridates considered here, which places these materials in a similar regime to the well-established $J_{\rm eff}$~$=$~1/2 magnet Sr$_2$IrO$_4$ ($\lambda/\Delta$~$>$~2.8)\cite{17_lu} and the Kitaev materials Na$_2$IrO$_3$ and $\alpha$-Li$_2$IrO$_3$ ($\lambda/\Delta$~$>$~4.7)\cite{13_gretarsson}. Taken together with the extremely small IrO$_6$ octahedral distortions determined by NPD and the enhanced branching ratio found by XAS, this is strong evidence that these double perovskite iridates host $J_{\rm eff}$~$=$~1/2 electronic ground states. 

Interestingly, although the intraband $t_{2g}$ crystal field excitations in Fig. 4 are very similar in all four double perovskite compounds studied (in terms of peak position, peak splitting, and linewidth), there are quite obvious differences in the properties of the interband $t_{2g}$ to $e_{g}$ transitions.  In particular, the $t_{2g}$ to $e_{g}$ excitations in Ba$_2$CeIrO$_6$ and Sr$_2$CeIrO$_6$ are much broader than those of La$_2$MgIrO$_6$ and La$_2$ZnIrO$_6$, indicating that the $e_{g}$ energy levels must possess a much larger electronic bandwidth. This difference is quite surprising, given that the intra-$t_{2g}$ excitations in the DP iridates appear to be largely insensitive to chemical composition or local structural details. At even higher energy transfers ($\hbar \omega $~$\sim$ 6 eV, not fully shown in Fig. 4), we observe a third set of transitions, which we attribute to charge transfer excitations from the O 2p band to the Ir 5d band. The energy and width of these charge transfer excitations appear to be almost identical in all compounds measured.

\section{V. Magnetic structures}

Now that we have established $J_{\rm eff}$~$=$~1/2 magnetism in La$_2$MgIrO$_6$, La$_2$ZnIrO$_6$, Ba$_2$CeIrO$_6$ and Sr$_2$CeIrO$_6$, it is fully anticipated that the classical phase diagram for the FCC Heisenberg-Kitaev-$\Gamma$ model\cite{15_cook} should serve as a good starting point for explaining their magnetic properties. Although La$_2$MgIrO$_6$ and La$_2$ZnIrO$_6$ have already been investigated with this in mind\cite{15_cook, 16_aczel}, it is important to note that a unique magnetic structure solution for La$_2$MgIrO$_6$ has not been found. While previous NPD work has identified A-type magnetic order in La$_2$MgIrO$_6$ and La$_2$ZnIrO$_6$\cite{13_cao}, this is consistent with both the A-I and A-II states that appear in the phase diagram. Both spin arrangements consist of FM planes that are stacked in an AFM fashion and ordered moments that point along Ir-O bonds, but the moments are perpendicular to the FM planes in the A-I state and parallel to the FM planes in the A-II state. Unfortunately, NPD cannot be used to determine the magnetic moment direction due to the observation of only a single magnetic Bragg peak at $Q$~$=$~0.79~\AA$^{-1}$ in each case.  It is important to distinguish between the A-I and A-II states experimentally because they establish the sign of a possible Kitaev interaction in these materials\cite{15_cook}. More specifically, an AFM (FM) Kitaev interaction is only compatible with the A-II (A-I) state. 

\begin{figure}
\centering
\scalebox{0.35}{\includegraphics{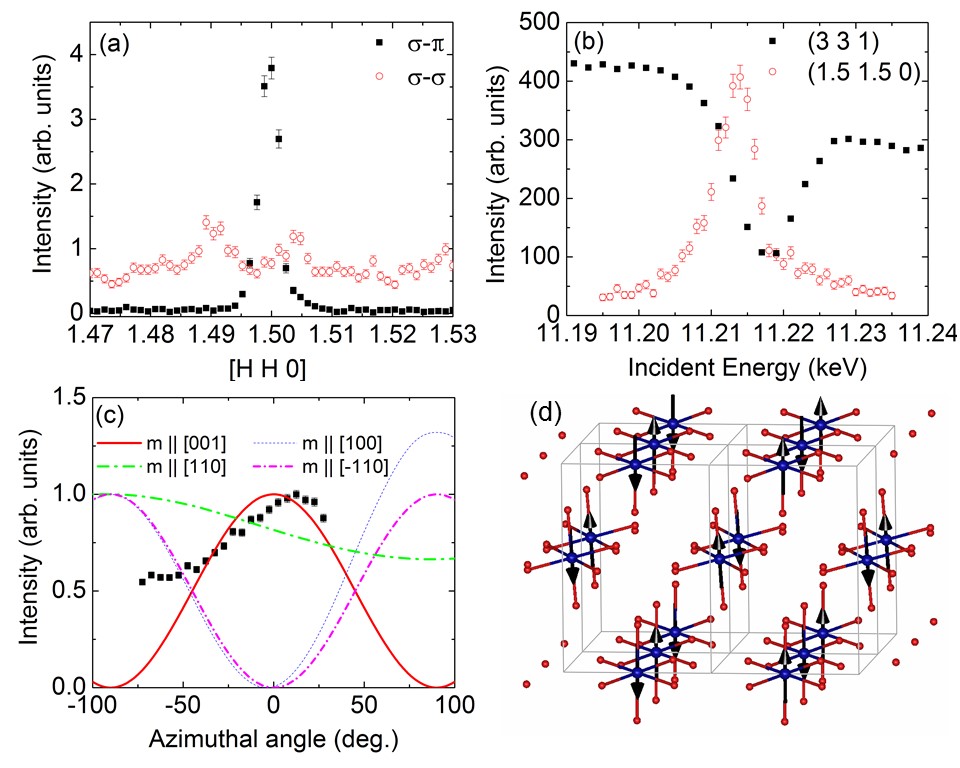}}
\caption{\label{Fig5} (color online) Resonant magnetic x-ray scattering results on single crystalline La$_2$MgIrO$_6$. (a) The structurally-forbidden Bragg peak (1.5 1.5 0) only appears in the $\sigma$-$\pi$ scattering channel. (b) The same (1.5 1.5 0) peak shows a resonant enhancement just below the Ir $L_3$ absorption edge, while the (3 3 1) structural peak exhibits typical intensity modulation near the absorption edge. (c) The azimuthal intensity dependence of the (0.5 0.5 0) Bragg peak is best described by A-II AFM order with the Ir moments aligned very close to the crystalline c-direction. (d) A schematic of the proposed magnetic structure for La$_2$MgIrO$_6$. It is anticipated that the Ir moments cant away from the c-axis to track the small IrO$_6$ octahedral rotations.  }
\end{figure}

\begin{figure*}
\centering
\scalebox{0.7}{\includegraphics{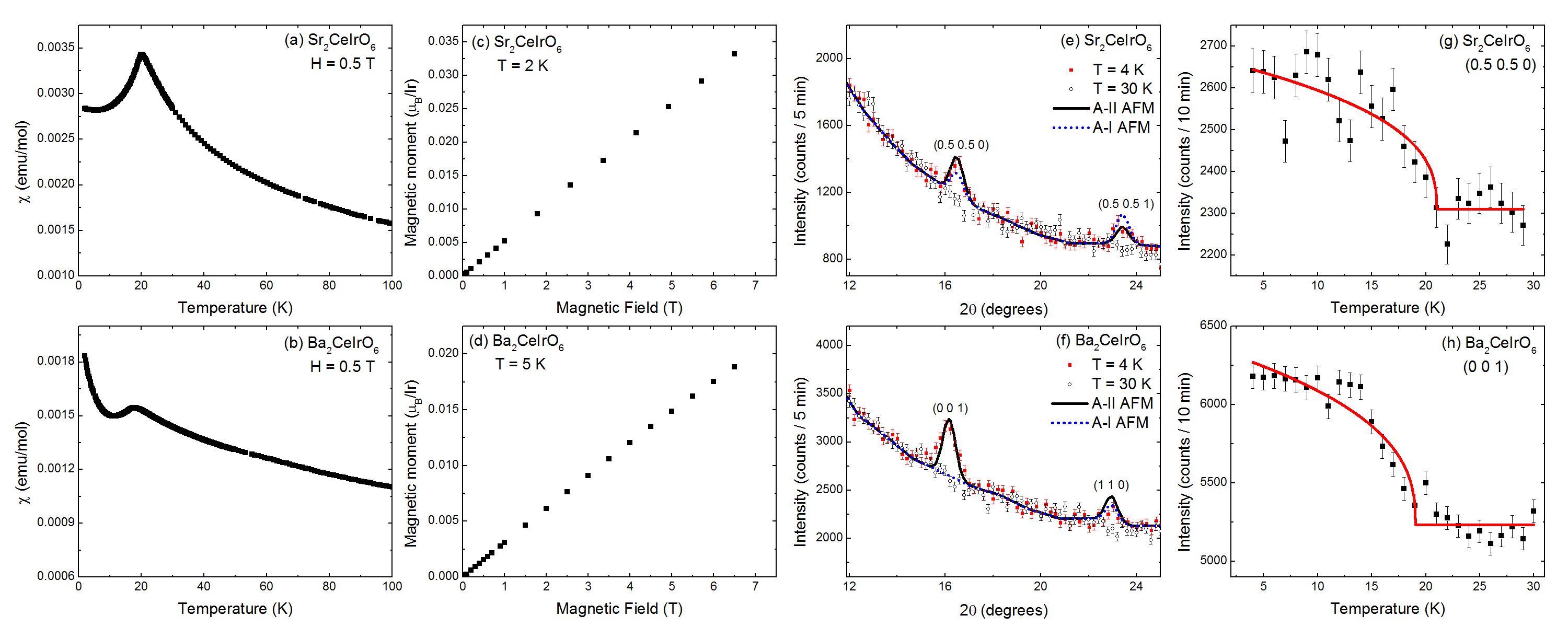}}
\caption{\label{Fig6} (color online) Bulk characterization and neutron powder diffraction (NPD) measurements on polycrystalline Sr$_2$CeIrO$_6$ and Ba$_2$CeIrO$_6$. (a), (b) Magnetic susceptibility measurements as a function of temperature reveal ordering temperatures $T_N$~$=$~21~K and 17~K for the Sr and Ba analogs respectively, which is in good agreement with previous work\cite{99_wakeshima, 99_harada}. (c), (d) The magnetization of both materials increases linearly with field. (e), (f) Neutron powder diffraction data is presented for both materials at $T$~$=$~4~K and 30~K. Two different refinement results are superimposed (solid and dotted curves) on the Sr$_2$CeIrO$_6$ and Ba$_2$CeIrO$_6$ data that are consistent with the field-dependence of the magnetization. The A-II AFM model provides superior agreement with the NPD data in both cases. (g), (h) Order parameter plots for the most intense magnetic Bragg peak observed in each case. The $T_N$ values are consistent with the magnetic susceptibility results. Power law curves are superimposed on the data as a guide to the eye.  }
\end{figure*}

In the absence of single crystals, some progress can be made towards establishing a unique magnetic structure for La$_2$ZnIrO$_6$ by assuming that the ordered moment direction tracks the IrO$_6$ octahedral rotations, as previously proposed\cite{09_jackeli} and subsequently verified for Sr$_2$IrO$_4$\cite{13_boseggia}. Although the single magnetic Bragg peak can be indexed in monoclinic notation with a magnetic propagation vector of either $\vec{k}$~$=$~0 or (0.5 0.5 0), which correspond to A-type AFM ordered states with FM plane stacking directions parallel and perpendicular to the c-axis respectively, the observation of a net FM moment in the magnetization vs magnetic field data\cite{13_cao} is only consistent with $\vec{k}$~$=$~0. The single magnetic Bragg peak must then correspond to $\vec{Q}$~$=$~(0 0 1) and its non-zero intensity is only compatible with an A-II structure due to neutron polarization factor arguments. On the other hand, similar reasoning cannot be used to differentiate between the A-I and A-II states for La$_2$MgIrO$_6$. The magnetization of this system increases linearly with magnetic field\cite{13_cao}. Since the IrO$_6$ octahedral rotations in these two materials are very similar as shown in Table~I, this implies that the magnetic propagation vector is $\vec{k}$~$=$~(0.5 0.5 0). However, both $\vec{Q}$~$=$~(0.5 0.5 0) or (0.5 -0.5 0) can contribute to the intensity of the single magnetic Bragg peak, which prevents straightforward differentiation between the A-I and A-II states. Single crystal measurements are therefore required in this case. 

Since we could not grow single crystals of La$_2$MgIrO$_6$ large enough for neutron diffraction, we performed a resonant magnetic x-ray scattering (RMXS) experiment to try and resolve this issue instead. With the photon energy tuned close to the $L_3$ edge and the sample cooled below $T_N$ to 6~K, new Bragg peaks emerged that are consistent with the expected magnetic propagation vector $\vec{k}$~$=$~(0.5 0.5 0). These extra peaks appear in the $\sigma$-$\pi$ channel only, as shown in Fig.~\ref{Fig5}(a) for the representative (1.5 1.5 0) position. We also plot the energy-dependence of this same peak and the structurally-allowed (3 3 1) reflection in Fig.~\ref{Fig5}(b). There is clear resonant behavior at the (1.5 1.5 0) position near the energy corresponding to the $L_3$ edge, while typical intensity modulation near the absorption edge is found at the (3 3 1) position. Finally, we find that the resonant enhancement of the (1.5 1.5 0) peak occurs slightly below the XAS maximum as noted for the magnetic peaks previously identified by RMXS in Sr$_2$IrO$_4$\cite{09_kim}, Na$_2$IrO$_3$\cite{11_liu}, and Sr$_3$Ir$_2$O$_7$\cite{12_boseggia}. These combined findings provide strong evidence that the (1.5 1.5 0) Bragg peak observed here has a magnetic origin.  

We proceeded to measure the azimuthal intensity dependence of the (0.5 0.5 0) magnetic Bragg peak in an effort to differentiate between the A-I and A-II states described above. This approach involves rotating the sample around the scattering vector $\vec{Q}$ in fixed increments and performing $\theta$ (i.e. rocking) scans at each of these different $\Psi$ angles. It has been used successfully to determine the ordered moment direction with RMXS in other antiferromagnetic materials\cite{13_boseggia_2}. Fig.~\ref{Fig5}(c) shows the integrated intensity of the (0.5 0.5 0) reflection as a function of $\Psi$ with solid black squares; $\Psi$~$=$~0 corresponds to the crystalline c-axis. The azimuthal intensity modulation expected for four different moment directions, as calculated by the software package FDMNES\cite{01_joly}, is also shown in this figure. The experimental data most closely resembles the calculation for ordered moments along the c-direction, which allows us to conclude that La$_2$MgIrO$_6$ has an A-II ground state. A schematic of the magnetic structure for La$_2$MgIrO$_6$ is shown in Fig.~\ref{Fig5}(d). 

Less information is known about the magnetic structures of Ba$_2$CeIrO$_6$ and Sr$_2$CeIrO$_6$. Previous bulk characterization studies have established AFM order in both materials with $T_N$~$=$~17~K and 21~K for the Ba\cite{99_wakeshima} and Sr\cite{99_harada, 00_harada, 16_karungo} analogs respectively, but the specific spin configurations have not been determined. We first present magnetic susceptibility vs temperature measurements in Fig.~\ref{Fig6}(a) and (b) for both materials, which show that the magnetic ordering temperatures of our samples are in good agreement with previous work. We also present low-temperature magnetization vs field data in Fig.~\ref{Fig6}(c) and (d), which shows the same linear behavior observed previously for La$_2$MgIrO$_6$ but not La$_2$ZnIrO$_6$. 

Since no magnetic Bragg peaks were observed in the HB-2A data, we collected complementary elastic neutron scattering data on the HB-1A triple-axis spectrometer. This instrument has an excellent signal-to-noise ratio and therefore is extremely useful for investigating materials with weak magnetic signals\cite{18_xiong}. Representative data indicated by the solid symbols and obtained at both 4~K and 30~K is shown in Fig.~\ref{Fig6}(e) for Sr$_2$CeIrO$_6$ and Fig.~\ref{Fig6}(f) for Ba$_2$CeIrO$_6$. Two new peaks are visible in the 4~K dataset of each material, and the order parameter plots presented in Fig.~\ref{Fig6}(g) and (h) indicate that they have a magnetic origin. To refine the magnetic structures with FullProf, we first fixed all of the structural parameters to the values obtained from the HB-2A refinements and then obtained an overall scale factor using the nuclear Bragg peaks measured with HB-1A (not shown here). 

Although the magnetic Bragg peaks of Sr$_2$CeIrO$_6$ can be indexed with either the A-type AFM propagation vector $\vec{k}$~$=$~0 or (0.5 0.5 0), the lack of a net FM moment despite the significant IrO$_6$ octahedral rotations is only consistent with the latter. The two magnetic Bragg peaks of Ba$_2$CeIrO$_6$ can also be attributed to A-type AFM, which has a propagation vector of $\vec{k}$~$=$~(0 0 1) for an ideal FCC structure. Therefore, we tried refinements with the appropriate propagation vector for each material using both A-I and A-II spin configurations. The Ir$^{4+}$ magnetic form factor, $j_0(Q)$, was taken from Ref.~\cite{11_kobayashi}. The results are superimposed on the data shown in Fig.~\ref{Fig6}(e) and (f) for the Sr and Ba systems respectively, and indicate that the relative intensities of the two magnetic peaks are captured much better by the A-II model in each case. Our best refinements yield ordered moments of 0.5(1)~$\mu_B$ for Sr$_2$CeIrO$_6$ and 0.3(1)~$\mu_B$ for Ba$_2$CeIrO$_6$. With A-II magnetic structures now established for both La$_2$MgIrO$_6$ and La$_2$ZnIrO$_6$, we also refined the ordered moment size for these materials using data from the neutron scattering experiments described in Ref.~\cite{13_cao}. We found an ordered moment of 0.6(1)~$\mu_B$ for La$_2$MgIrO$_6$ and an ordered moment of 0.3(1)~$\mu_B$ for La$_2$ZnIrO$_6$. The common A-II AFM order in these four materials is likely driven by a significant AFM Kitaev interaction.  

\section{VI. Magnetic Hamiltonians}

Perhaps the most interesting aspect of La$_2$MgIrO$_6$ and La$_2$ZnIrO$_6$ is that their thermodynamic properties, A-II ordered states, and gapped magnetic excitation spectra can be explained by a Heisenberg-Kitaev Hamiltonian with a dominant AFM Kitaev term\cite{15_cook, 16_aczel}. Unfortunately, the monoclinic structural distortions of these materials ensure that competing Heisenberg-Ising models cannot be ruled out based on symmetry grounds alone and different spin gap origins become viable within the Heisenberg-Kitaev models\cite{16_aczel}. Since a common microscopic mechanism appears to be driving the magnetism of Ba$_2$CeIrO$_6$, Sr$_2$CeIrO$_6$, La$_2$MgIrO$_6$, La$_2$ZnIrO$_6$, as indicated by the realization of $J_{\rm eff}$~$=$~1/2 moments and A-II AFM order in all four materials, a systematic study of their spin waves may allow one to place more stringent constraints on possible model Hamiltonians and spin gap mechanisms. 

\begin{figure}
\centering
\scalebox{0.35}{\includegraphics{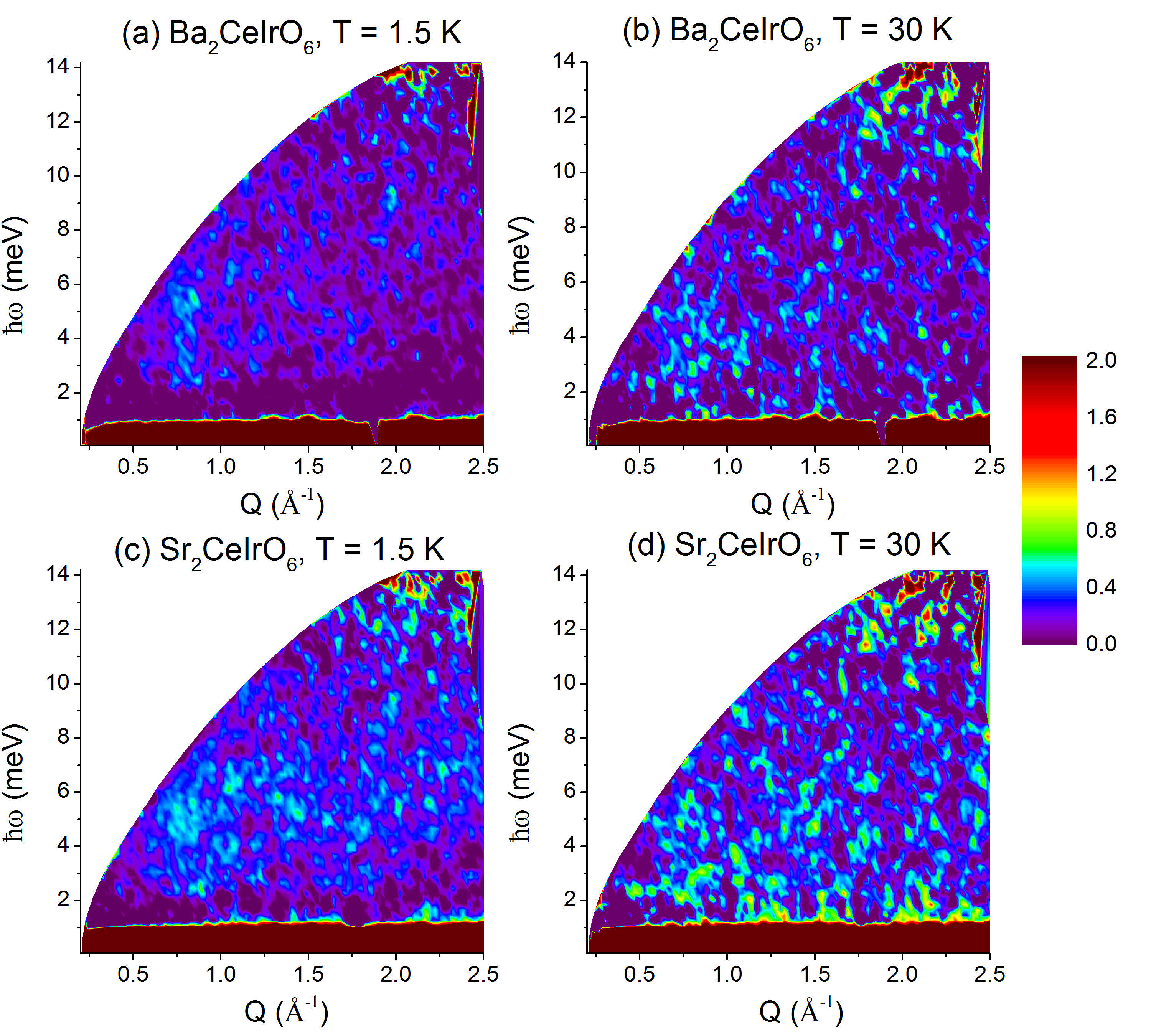}}
\caption{\label{Fig7} (color online) Color contour plots of the $E_i$~$=$~15~meV HYSPEC data for Ba$_2$CeIrO$_6$ at (a) $T$~$=$~1.5~K and (b) $T$~$=$~30~K. Similar plots are shown in (c) and (d) for Sr$_2$CeIrO$_6$. In sharp contrast to previous work on La$_2$MgIrO$_6$ and La$_2$ZnIrO$_6$\cite{16_aczel}, no clear magnetic excitations are visible in these spectra. }
\end{figure}

Previously, we had great success measuring the magnetic excitation spectra of polycrystalline La$_2$MgIrO$_6$ and La$_2$ZnIrO$_6$ by performing an inelastic neutron scattering experiment at the HYSPEC spectrometer\cite{16_aczel}. Therefore, we used the same experimental set-up to investigate the spin dynamics of Ba$_2$CeIrO$_6$ and Sr$_2$CeIrO$_6$. Our main results are summarized in Fig.~\ref{Fig7}; color contour plots for Ba$_2$CeIrO$_6$ and Sr$_2$CeIrO$_6$ are presented with an incident energy $E_i$~$=$~15~meV and temperatures both above and below $T_N$. Inelastic spectra were also collected for both materials at the same temperatures with $E_i$~$=$~7.5~meV (not shown). Surprisingly, in sharp contrast to our previous work on La$_2$MgIrO$_6$ and La$_2$ZnIrO$_6$, no clear magnetic signal is observed in this data, thus precluding detailed comparisons of the spin dynamics of these four materials. This difference in the inelastic spectra of the Ce and La samples is observed despite their similar ordered moments. The higher magnetic ordering temperatures of Ba$_2$CeIrO$_6$ and Sr$_2$CeIrO$_6$, as compared to La$_2$MgIrO$_6$ and La$_2$ZnIrO$_6$, imply larger exchange interactions in the Ce samples and hence an increased spin wave bandwidth. As these excitations become more dispersive, they will be harder to detect for a comparable signal-to-noise ratio and this may be the reason that they were not observed in the Ba$_2$CeIrO$_6$ and Sr$_2$CeIrO$_6$ spectra. Single crystal measurements are therefore required to measure the spin waves in the Ce samples with INS.

\section{VII. Conclusions}
By providing strong experimental evidence for Ir$^{4+}$ $J_{\rm eff}$~$=$~1/2 electronic ground states in the double perovskite iridates La$_2$MgIrO$_6$, La$_2$ZnIrO$_6$, Ba$_2$CeIrO$_6$ and Sr$_2$CeIrO$_6$, we showed that the classical phase diagram for the FCC Heisenberg-Kitaev-$\Gamma$ model discussed in Ref.~\cite{15_cook} should provide an excellent starting point for explaining their collective magnetic properties. We strengthened this conjecture by identifying magnetic structures in these materials that appear on this phase diagram. More specifically, our scattering results on La$_2$MgIrO$_6$, Sr$_2$CeIrO$_6$, and Ba$_2$CeIrO$_6$ establish A-II AFM order in these materials, which was previously identified for La$_2$ZnIrO$_6$\cite{16_aczel}. We anticipate that this magnetic ground state arises from a significant AFM Kitaev interaction that is common to all four of these double perovskite iridates. 

Attempts to determine the magnetic Hamiltonian for Ba$_2$CeIrO$_6$ and Sr$_2$CeIrO$_6$ with inelastic neutron scattering and therefore prove this hypothesis were unsuccessful, as no magnetic excitations were observed in this data. Nonetheless, our combined results reported here elucidate the extreme similarities in the electronic and magnetic properties of these four systems and suggest that spacing $J_{\rm eff}$~$=$~1/2 magnetic atoms further apart is a promising way to find new Kitaev material candidates with lattice geometries beyond honeycomb. The FCC magnets $A_2$IrX$_6$ where $A$~$=$~Na, K, Rb, or Cs and $X$~$=$~F\cite{17_rossi} or Cl\cite{67_hutchings, 76_lynn} are particularly attractive in this regard, as adjacent IrX$_6$ octahedra are arranged in the appropriate geometry required to realize significant Kitaev interactions through extended superexchange pathways and ideal cubic systems may exist\cite{67_hutchings, 76_lynn}. 

{\it Note Added:} While finishing up this manuscript, we noticed that a related study on Ba$_2$CeIrO$_6$ was posted on the preprint server \cite{19_revelli}. In that work, the authors report single crystal x-ray diffraction data that is well described by the {\it Fm$\bar{3}$m} space group and RIXS measurements of the intraband $t_{2g}$ crystal field excitations that support a $J_{\rm eff}$~$=$~1/2 electronic ground state. They also perform theoretical calculations to show that the magnetic Hamiltonian for Ba$_2$CeIrO$_6$ consists of a significant AFM Kitaev interaction. These results are all consistent with the conclusions reported in our manuscript. The main difference between the two works is we find an experimental magnetic propagation vector $\vec{k}$~$=$~(0 0 1), while the theoretical calculations in Ref.~\cite{19_revelli} predict $\vec{k}$~$=$~(0 0.5 1).  

\section{acknowledgments}

Materials synthesis and characterization at ORNL were supported by the U.S. Department of Energy, Office of Science, Basic Energy Sciences, Materials Sciences and Engineering Division (J.-Q.Y.).  A portion of this research used resources at the High Flux Isotope Reactor, which is a DOE Office of Science User Facilities operated by Oak Ridge National Laboratory. This research used resources of the Advanced Photon Source, a US Department of Energy (DOE) Office of Science User Facility operated for the DOE Office of Science by Argonne National Laboratory under contract No. DE-AC02-06CH11357. Research conducted at the Cornell High Energy Synchrotron Source (CHESS) is supported by the NSF \& NIH/NIGMS via NSF award DMR-1332208. Research at the University of Tennessee and the University of Illinois is supported by the National Science Foundation, Division of Materials Research under awards DMR-1455264 (G.J.M) and DMR-1350002 (H.D.Z).

\end{document}